\documentclass[]{aiaa-tc}
\usepackage{graphicx}
\usepackage[verbose]{wrapfig}
\usepackage{amsmath}
\usepackage[hidelinks]{hyperref}

 \title{Fully Optical Spacecraft Communications: Implementing an Omnidirectional PV-Cell Receiver and 8Mb/s LED Visible Light Downlink with Deep Learning Error Correction}

 \author{
  Sihao Huang and Haowen Lin\\
  {\normalsize
	Aphelion Orbitals, Inc.
}\\
\emph{3433 Joe Murell Dr, Titusville FL 32780}
}

 \AIAApapernumber{YEAR-NUMBER}
 \AIAAconference{Conference Name, Date, and Location}
 \AIAAcopyright{\AIAAcopyrightD{YEAR}}

\begin{document}

\maketitle

\begin{abstract}
Free space optical communication techniques have been the subject of numerous investigations in recent years, with multiple missions expected to fly in the near future. Existing methods require high pointing accuracies, drastically driving up overall system cost. Recent developments in LED-based visible light communication (VLC) and past in-orbit experiments have convinced us that the technology has reached a critical level of maturity. On these premises, we propose a new optical communication system utilizing a VLC downlink and a high throughput, omnidirectional photovoltaic cell receiver system. By performing error-correction via deep learning methods and by utilizing phase-delay interference, the system is able to deliver data rates that match those of traditional laser-based solutions. A prototype of the proposed system has been constructed, demonstrating the scheme to be a feasible alternative to laser-based methods. This creates an opportunity for the full scale development of optical communication techniques on small spacecraft as a backup telemetry beacon or as a high throughput link.
\end{abstract}

\section{Introduction}
The large surface area of the PV cells on a spacecraft provides a significant sensor area which can be utilized as an omnidirectional receiver on many solar panel layouts. Previous researchers have proposed the use of this unique asset as a backup communication system\cite{GuoAndThangavelautham}, realizing that a low cost, low power consumption uplink could be created by using the photovoltaics as a laser receiver. Such proposed systems have data rates of less than 10Kb/s, allowing them to serve as effective emergency alternative-band communication systems, though not as an operational uplink.
 


PV cells are intrinsically difficult to work with as high-bandwidth components due to their parasitic capacitance and high photocarrier drift and diffusion times. Building on the work of Won-ho Shin et al.\cite{Shin:16}, which demonstrated 17.05 Mbps discrete multitone transmission on a solar panel receiver, we propose an implementation of overcoming these limitations through reverse biasing with little impact on the overall efficiency of the power system.

One of the most significant advantages of the use of a PV-cell receiver is the low mass and power penalty to a small spacecraft. This becomes more apparent on Cubesats and even femtosatellites, negating the need for an uplink receiver or antenna, providing mass, cost, and power budget savings in a design environment where all three are critical to mission success.
 
Successful commercial implementation of a fully optical communication system depends on high reliability and a low requirement for pointing accuracy. The second part of the link utilizes a non-coherent, LED-based transmitter which enables a stable downlink without a high accuracy attitude determination and control system (ADCS). The use of LED beacons is not new; the idea has been trialled on FITSAT-1, which successfully recovered a 5kHz modulation on the signal\cite{niwakaSat}. ShindaiSat \cite{ShindaiSat} flown in 2014 utilized an array of 36 LEDs on its high gain array, with a total output of 86.4W. This allowed for a data rate of 9.6Kb/s while utilizing basic FSK modulation and white light, which negated the possibility of spectrum filtering. Moreover Li-Fi and similar technologies have matured the field of visible light communication considerably and the characteristics of LEDs are well known\cite{lifiBook}.  


 
\begin{figure}
\begin{minipage}{0.5\textwidth}

\includegraphics[width=\textwidth, trim={0 0 0 14pt}, clip=true]{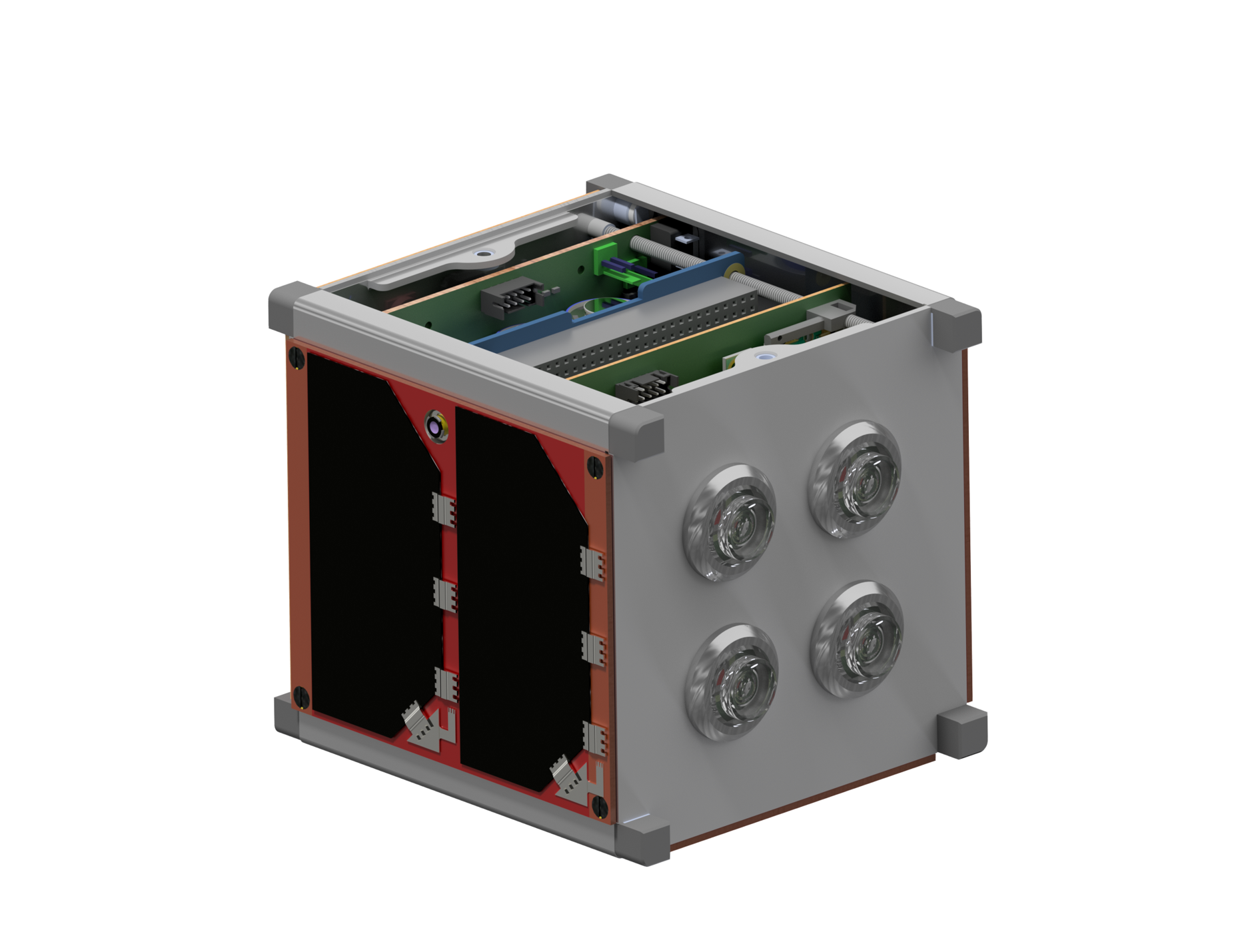}

\caption{Implementation of proposed VLC transmitter and PV-cell receiver on the Calypso Spacecraft}
\end{minipage}
\begin{minipage}{0.55\textwidth}

\includegraphics[width=\textwidth, scale=1.2]{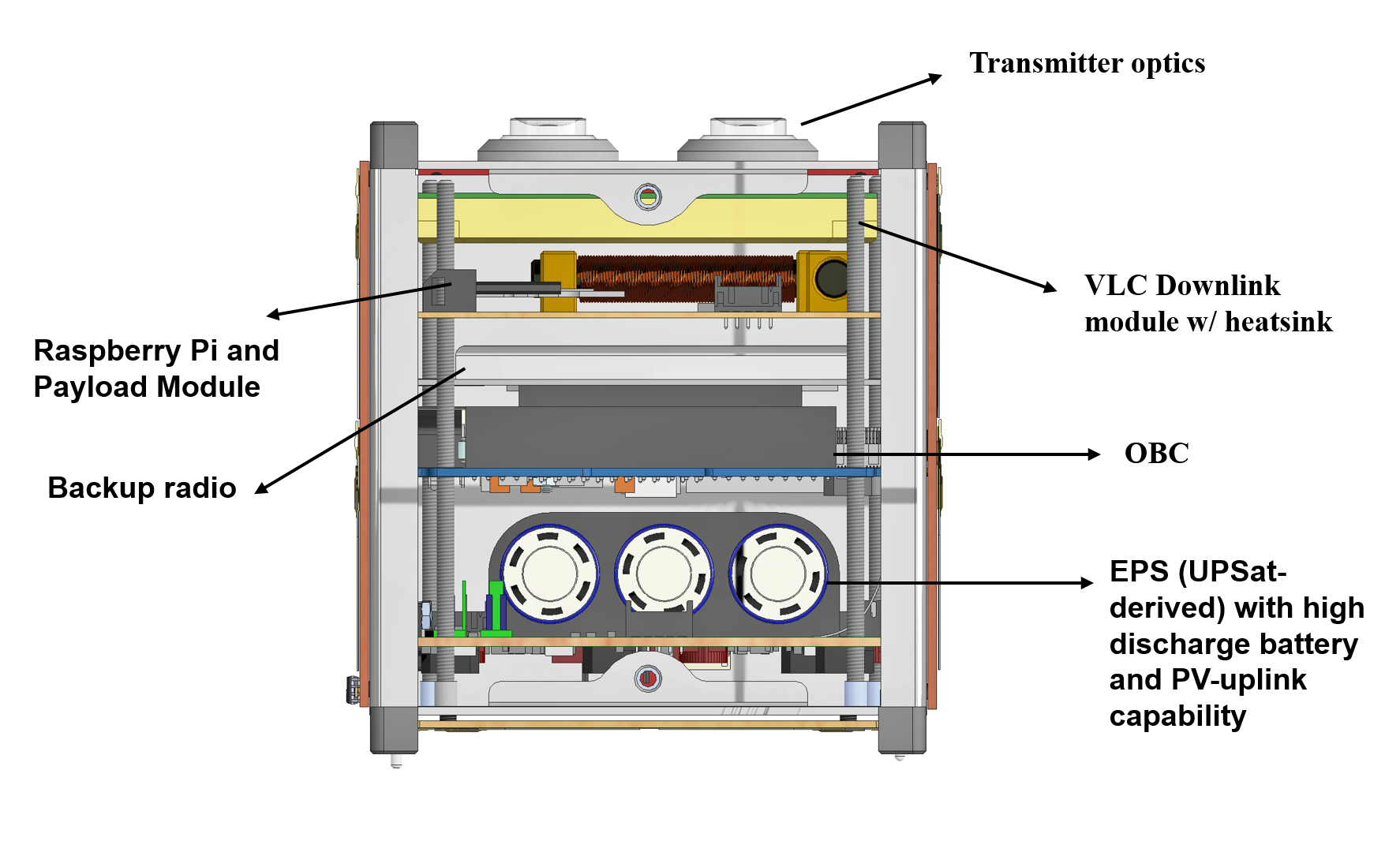}

\vspace{+7pt}
\caption{Internal layout of the Calypso 1U Cubesat}
\end{minipage}

\end{figure}

Despite lower directionality, the significant efficiency improvement of LED transmitters (state-of-the-art at around 70\% for LEDs, 30\% for laser diodes) enables a power-per-bit value that is much lower than expected. The data rate ceiling for a transmitter bound by the power and pointing limits of a 3U Cubesat is predicted to be around 50Mb/s. This, too, sparked significant internal interest for further research as it indicates a large potential in nanosatellite applications to fill the gap in communication bandwidth. Future advancements such as larger ground stations, improved focusing systems, and optimized modulation/recovery techniques will enable the envelope to be pushed even further.

\bigskip

Amongst the biggest challenges in implementing a commercial VLC system are:
 
\begin{itemize}
\item The hardware implementation and the miniaturization of a high power, high bandwidth system
\item The difficulties presented by atmospheric distortion and the compensation techniques necessary to counteract these impairments
\item The relatively poor signal to noise ratio of non-coherent carriers compared to laser-based systems
\end{itemize}
 
In this paper we address all three issues respectively.





\section{Link Design}

Signal to noise ratio is a prime concern on the spacecraft downlink due to the low monochromaticity of single-color LEDs (most commercial-off-the-shelf components have a spectral spread of around 15nm). At the same time, the design is restricted by a wider beam. Nevertheless LEDs allow for higher power density for its footprint compared to laser-based systems. To this we have chosen a peak electrical power of 140W and average operational power of 15W, with an electro-optical efficiency of 70\%, well within the feasible limits of a 1U Cubesat. 

The gain of the transmitter and receiver are given by the following\cite{FreeSpaceLaserCommunications}: \[G(tx)=\frac{16d^2}{\lambda^2}\] 
\hspace{4pt} 
\[G(rx) = (\frac{\pi d}{\lambda})^2 \]

A 400 KM circular orbit is used with a maximum angle from zenith of 40°, a beam divergence of 20°, and a homogeneous power distribution. Note that as a result of the relatively large beam width, we approximate the aiming loss to 1dB resulting from receiver tracking inaccuracies. A 30cm (12 inch) commercially available telescope is used as the receiver and 700 photons/bit is assumed.

\begin{center}
\begin{tabular}{c c}
\hline
Transmit Power & 19.9 dBW\\ \hline
Transmitter Gain & 12 dBi\\ \hline
Path Length & 522 km\\ \hline
Free Space Path Loss & 260.5 dB\\ \hline
Atmospheric Loss & 2 dB \\ \hline
Pointing Loss & 1 dB \\ \hline
Receiving Aperture & 30 cm \\ \hline
Receiver Gain & 123.6 dBi \\ \hline
Received Power & -107.3 dBW \\ \hline
Bit Rate $(700\ photons/bit)$ & $8.85\cdot 10^4 bits/s$ \\ \hline
\end{tabular}
\end{center}

This is comparable to most high-performance S-band systems available on the market today and offers a similar bit rate to laser communication systems on the same scale. Overall the system offers a very attractive alternative to either solution due to its low cost and reduced ADCS requirements. It is also worth noting that with a radiant flux of $8.16 \cdot 10^{-9} W/m^2$ , the satellite will have a predicted visual magnitude of around 1, making it visible under good seeing conditions.
 
The uplink laser will utilize a 150W YAG-pumped 1064 nm laser mounted coaxially to the downlink receiver and tracking setup via a 0.3m aperture transmitter. The ground station cost is of a lower concern as such systems do not need to be made available to every spacecraft operator. The implementation of a pixel array-based tracking system allows for a tracking error of less than 1.2 arcsecond \((3 \sigma)\) without a fine steering mirror, and for spacecraft identification in emergencies where the exact orbit may not be known. A simple radiometric link budget gives the received power of the link.

\begin{center}
\begin{tabular}{c c}
\hline
Transmit Power & 21.76 dBW\\ \hline
Transmitter Gain & 121.0 dBi\\ \hline
Path Length & 522 km\\ \hline
Free Space Path Loss & 255.8 dB\\ \hline
Atmospheric Loss & 2 dB \\ \hline
Pointing Loss & 1 dB \\ \hline
Receiving Area & \(70 cm^2\) \\ \hline
Receiver Gain & 108.9 dBi \\ \hline
Received Power & -8.12 dBW \\ \hline
\end{tabular}
\end{center}

At a background irradiance of \(1000W/m^2\) (worse-case scenario), with a 8.12 dbW, or 0.154W, of received power, it is predicted that the peak-to-peak voltage of the signal is 0.45V. This allows for relatively easy detection using the electrical power system or a dedicated channel. The data ceiling of the system is thus bound mostly by the effective bandwidth of the solar cells.

\section{Implementation}
The proposed communication architecture, composed of a VLC downlink and PV-cell uplink, will be flown on Calypso, a 1U Cubesat developed jointly between Aphelion Orbitals and the Aerospace Research \& Engineering Systems Institute. The two functions are contained in a dedicated VLC downlink module and a custom EPS-integrated uplink driver. This mission serves to both demonstrate the low volume and ADCS requirements of the system by implementing it in the given form factor and power budget.

\subsection{Cell Based Uplink}
Reverse biasing photodiodes is a common technique for improving bandwidth by increasing the number of photocarriers and improving drift velocity. The use of self-biasing on solar cells has been investigated in the VLC industry\cite{Shin:16}. This technique has been demonstrated to improve the -3dB bandwidth of a PV cell by up to 60\% utilizing a 30V bias provided by a lightweight upconverter. Moreover, it was determined that minimal low energy losses were incurred as biasing recovers significant energy expenditure through increasing PV cell efficiency.
 
\begin{figure}[h]
\vspace{-7pt}
\includegraphics[width=\textwidth]{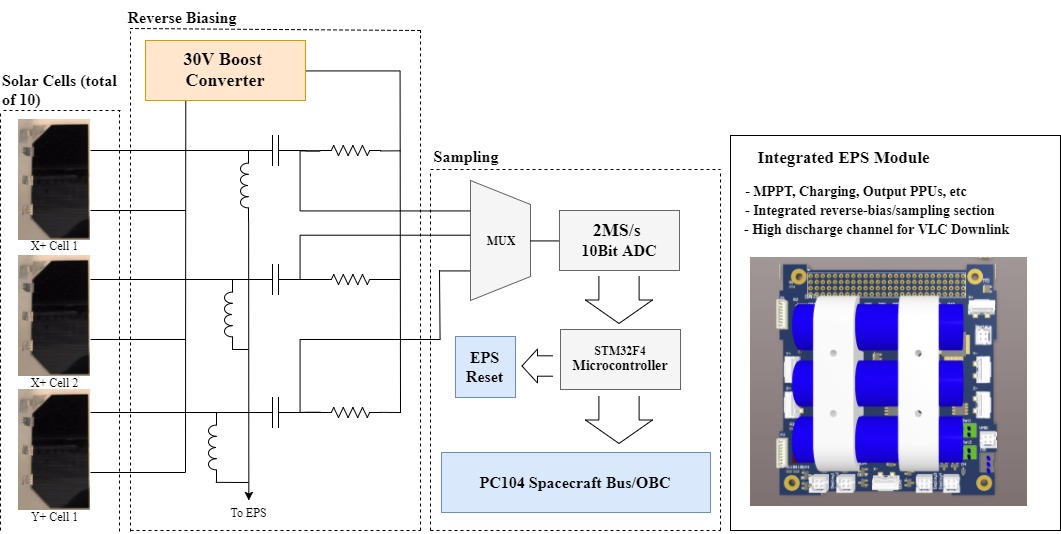}
\caption{Topology of the PV-cell uplink with internal reverse-biasing. The spacecraft EPS board performs all functions in an integrated module. Note that only 3 cells out of 10 are depicted.}
\end{figure}

The use of the uplink as the primary communication system requires that the receiver is powered on at all times. Thus we employed a low power wake-up scheme utilizing an envelope detector and a 2MS/s COTS ADC which polls all faces for a period of 100mS each at a reduced sampling rate. This scanning process can be implemented to consume minimal standby power as demonstrated by similar implementations in commercial wake-up receivers\cite{WakeUpSystem}. The ground station broadcasts a link start signal for 5 seconds. When a clock signal is recovered, the ADC selects the face with the best SNR and processes telemetry data.
 
One of the primary goals of this payload is to investigate the feasibility of such a system as an emergency communication and reset channel. The implementation of the receiver circuitry in the EPS allows it to power cycle the spacecraft via dedicated control. At the same time, the received telemetry can be delivered to the OBC to overrides radio link data.

\subsection{Visible Light Downlink} 

A survey of COTS LEDs has revealed that, though limited, there exists a number of options available that provide built-in lensing at a 20-degree beam width. However, to open up the possibility for more efficient, monolithically packaged, high power diodes, additively manufactured, low profile lenses were selected. They can be used without major thermal concern due to the low duty-cycle of the transmitter. The implementation on Calypso utilizes four high efficiency, high power, 620 nm Luminus SBT-90 LEDs capable of 1600 lumens at 13.5A peak power each.

To maximize transmit antenna gain, we have evaluated a number of commercial TIR lens and parabolic reflector attachments. As no suitable lightweight component was found for the large aperture of the Luminus LEDs, a set of additively manufactured lenses based on an SLA process will be produced and subsequently processed for a suitable surface finish.

\begin{wrapfigure}[8]{r}{0.3\textwidth}
\vspace{-12pt}

\includegraphics[width=0.3\textwidth]{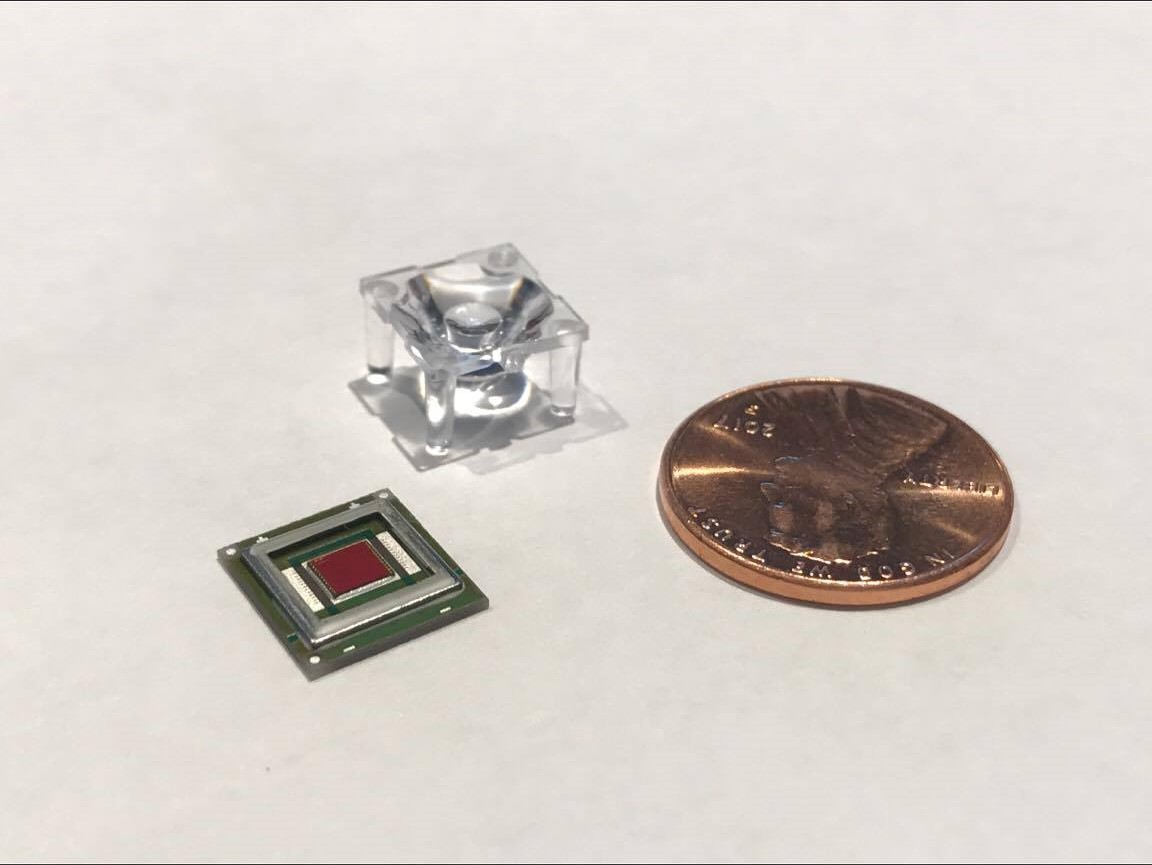}

\caption{A Luminus SBT-90 LED with an evaluated COTS TIR lens}

\end{wrapfigure}

The Calypso payload implements a downlink LED driver with 140W peak power and a 40Mb/s bandwidth limit. Scaling down a transmitter of this power represented a sizable technical obstacle due to thermal dissipation.

This was mitigated through the low specified duty cycle which restricts the transmitter to a 2 minute on-time per orbit. An Altera Cyclone IV handles the stream processing and interfacing with the OBC. The FPGA, together with a Raspberry Pi compute module, implements an end-to-end deep learning-based signal processing system that drives a 125Ms/s, 14 bit DAC. This produces a QAM-16 modulated signal that is fed into four different high-powered bias-tees after amplification. A dedicated high discharge battery in the EPS feeds the synchronized pulsed MOSFET LED driver.

\begin{figure}[h]
\vspace{-12pt}

\includegraphics[width=\textwidth]{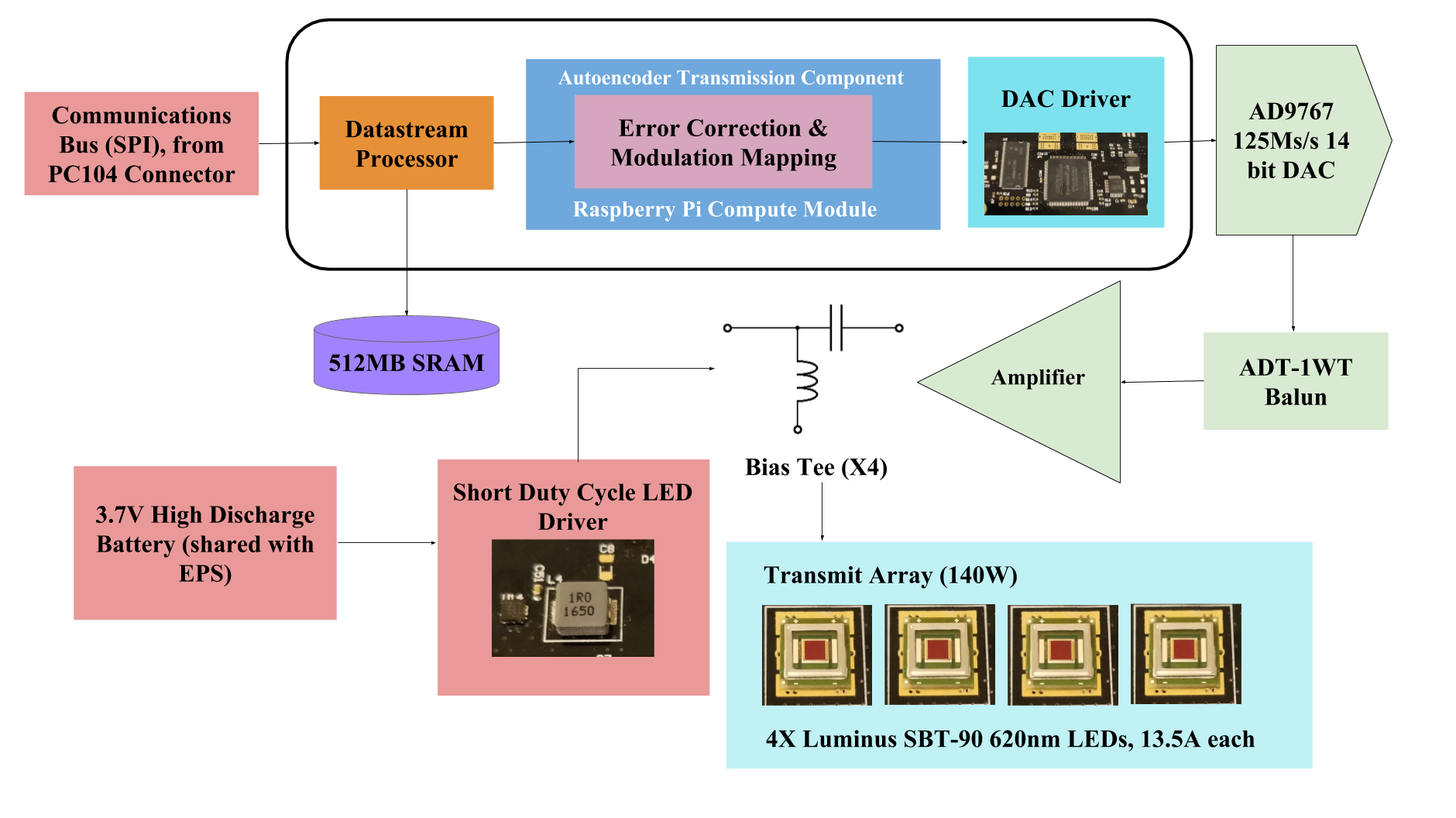}

\caption{High level block diagram of the VLC downlink transmitter}

\end{figure}
 
The module is contained within a single PC104 form-factor PCB mounted on the Z+ face of the Cubesat, weighing 150g. An integrated, 7075 aluminum housing on the back of the unit is directly coupled to the LEDs and switching circuitry to act both as an EMI shield and a heat sink. Prototypes of the driver circuitry have been constructed, as shown in Figure 5.
 
The units have been tested rigorously in Aphelion's facilities for power characteristics, modulation response, and thermal dissipation under vacuum conditions. A second revision will be made in accordance with the testing, which called for a higher powered preamplifier.

\begin{figure}
\begin{minipage}{0.5\textwidth}

\includegraphics[width=\textwidth, trim={0 0 0 14pt}, clip=true]{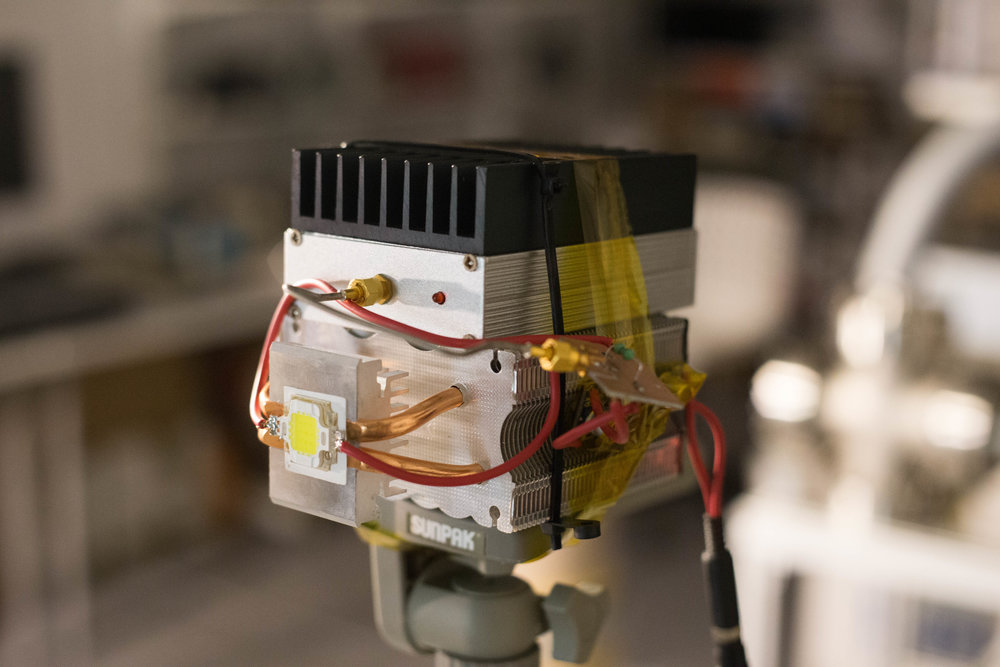}

\caption{Prototype VLC transmitter with amplifier and bias-t}
\end{minipage}
\begin{minipage}{0.5\textwidth}

\includegraphics[width=\textwidth]{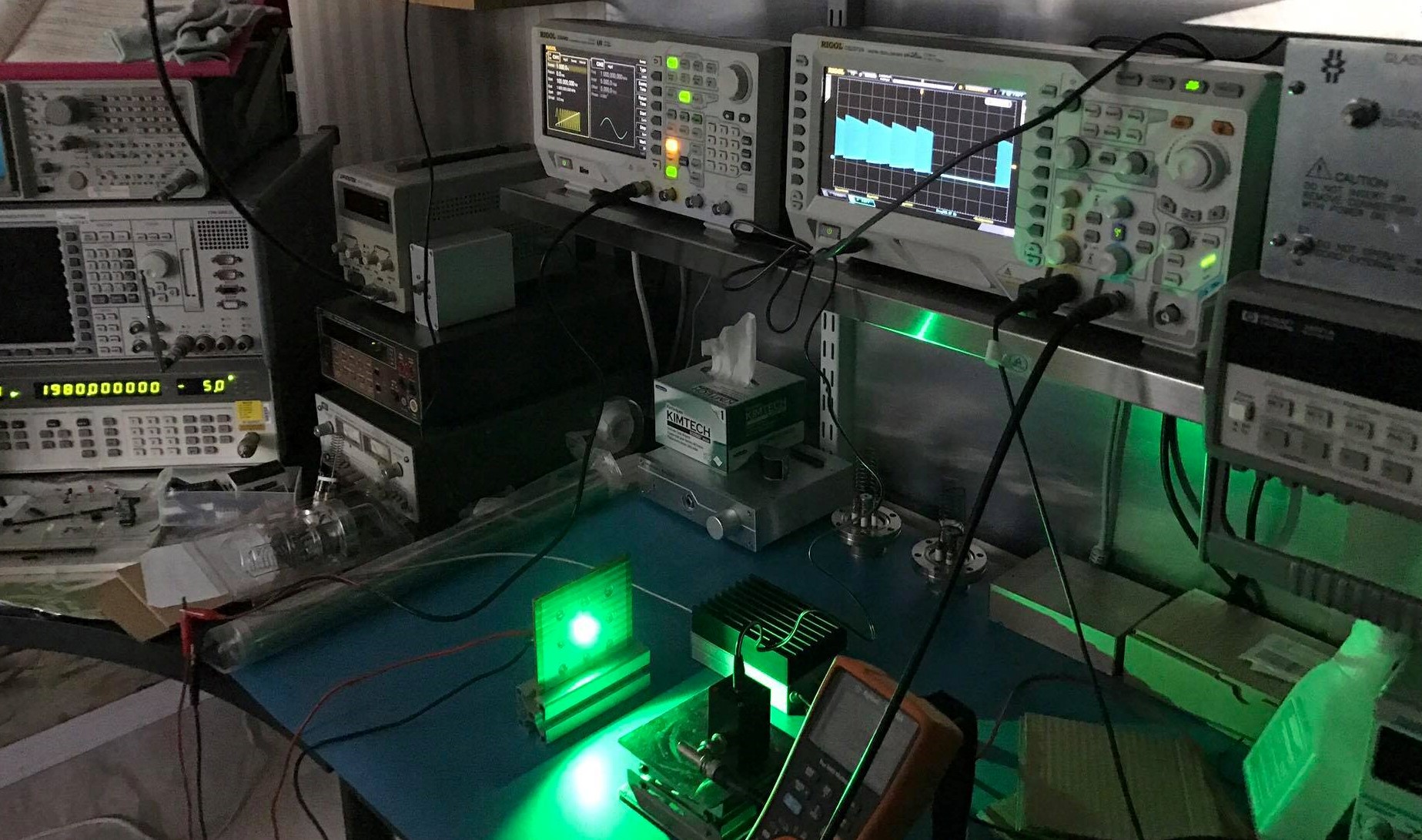}

\caption{Early transmitter test showing an FM sweep. A DC-biased QPSK link was demonstrated using the same hardware.}
\end{minipage}

\end{figure}







\section{Ground Station/Compensation}
Free space optical communication systems need to compensate for channel impairments such as atmospheric distortion and the frequency response of hardware components (i.e. transmitter LEDs). In order to account for such problems through an economical and lightweight method, an innovative method using deep learning was devised. Utilizing the research of O'Shea and Hoydis, we modeled the downlink as an autoencoder\cite{OSheaAndHoydis}, an unsupervised learning device used to map high dimensional data to lower dimensions, acting similar to a compression algorithm. This allows us to maximize the channel bandwidth. Communications systems generally require mathematically tractable models\cite{OSheaAndHoydis}. However, in the interest of lowering costs of development and deployment in production environments, it is more economically feasible to eschew traditional multiple stage communication systems in the favor of a single end-to-end deep learning system optimized for the specific hardware and channels, lowering development time and cost.
 
As certain types of deep learning systems (specifically, recurrent neural networks) are known to be Turing-complete\cite{Siegelmann} and universal function approximators\cite{Hornik}, we are able to combine the various parts of classical communication architecture into a single block and have the neural network model it in its entirety. The cost of deep learning systems have been falling rapidly and as efficiency of GPU systems improve, it is possible to embed such a system into a Cubesat. There are current research efforts in creation of specialized ASICs (Application Specific Integrated Circuits) optimized for neural network systems, including Google's Tensor Processing Units. In future iterations, it is possible to integrate such chips into Cubesats instead of off-the-shelf graphical processing units designed for embedded systems (in this case the Raspberry Pi on-board GPU) that will be flown on Calypso, allowing higher power efficiency and better optimization.
 
\begin{figure}[h]
\vspace{-8pt}
\includegraphics[width=\textwidth]{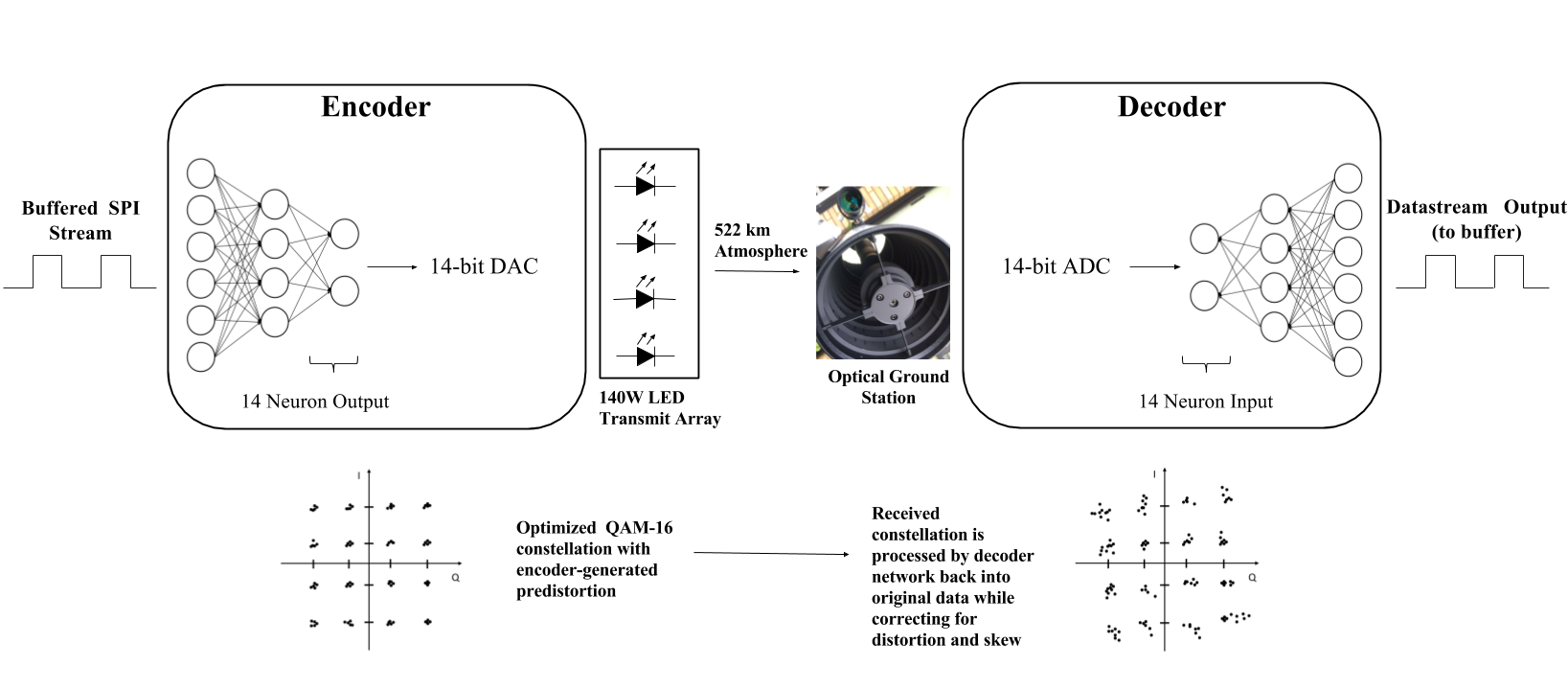}
\caption{Prototype VLC transmitter with amplifier and bias-t}
\end{figure}

The autoencoder system will implemented with a one-hot input vector of the maximum size the Calypso's embedded Raspberry Pi module is capable of efficiently running in flight conditions (which will be determined later empirically). Assuming an input vector of size $n$ where $ n = 16$, each packet of data fed into the encoder would be $log_2 n$ and in this case 4 bits of data. The goal of the neural network is to minimize the difference between the theoretically achievable SNR and the actual SNR. The theoretical transmission we are modeling upon utilizes a QAM-16 transmission system with Hamming(7,4) error correction. It has already been demonstrated by O'Shea and Hoydis that such machine learning based signal processing systems are capable of performing just as well if not exceeding the efficiency of a traditional encoding and/or modulation system that operate in discrete stages. On this premise, we have decided to train the neural network on Calypso while it is in flight instead of doing so on the ground with theoretical atmospheric loss models so as to achieve the highest efficiency in practice. The training data will be sent up via the main RF uplink and it will be used to train the optical communication system. The system proposed by O'Shea and Hoydis is implemented with the final  layer of the encoder being 7 neurons, due to the choice of error correction. The output of the encoder is passed through a I-Q modulator via the Altera before being pushed into DAC. Similarly, the ADC on the downlink also has its output passed through a I-Q modulator before before being fed into the decoder system. 

The use of machine learning as a replacement for a traditional signal processing stack offers great flexibility as the communication system can now be updated in-flight. The self-optimizing nature also means that it would attempt to attain the best performance possible with minimum human intervention.  

At the same time, we recognize that VLC offers an exciting opportunity in atmospheric correction technology.
When multiple transmitters are used on a single link, a differential phase delay can be observed, as recognized by the work of Yu Si Yuan\cite{PATBook}, which we summarize as follows. 

\begin{align*}
\phi_1 = k\int_0^H[n_0 - n_1 (z)]dz\\
\phi_2 = k\int_0^H[n_0 - n_1' (z)]dz
\end{align*}

Where $n_0$ is the average diffraction index, $n_1$ represents the fluctuation of that value, and k is the wavenumber of the transmitting beam. Hence, the variance of the phase delay can be stated as follows.

\[\sigma^2_\phi (H, \theta)=\langle(\phi_1-\phi_2)^2 \rangle =2.914k^2(\frac{d}{H})^{\frac{5}{3}} sec \hspace{2pt}\psi \int^4_0 C^2_n(z)z^{\frac{5}{3}} dz\] 

$C^2_n$, the atmospheric structure constant, is a function of height $z$ and is given by\\ 
$C^2_h (h) = 0.0059(\frac{v}{27})^2(10^{-5h})^{10} e^{-\frac{h}{1000}} + C_0e^{-\frac{h}{100}}$, $C_0$ is approximately $1.7 \cdot 10^{-14}$ .\\

It has been suggested that provided a large enough separation between the transmitters (in the order of multiple kilometers), the interference between the two beams can effectively correct for atmospheric scintillation. However, the differential phase delay is too small for use in atmospheric fluctuation correction on most spacecraft due to physical restraints preventing the two transmitters from being positioned a substantial distance apart.

However, the use of visible light communication techniques opens up the possibility of multiple synchronized receivers due to the wide beam coverage. The Calypso payload will evaluate the use of this technique using two identical Rubidium time-synchronized receivers placed 20KM apart to record the differential delay between the two received signals. The data is then interfered and corrected digitally to recover the transmission.

\begin{wrapfigure}[16]{r}{0.5\textwidth}	
\vspace{-10pt}

\includegraphics[width=0.5\textwidth]{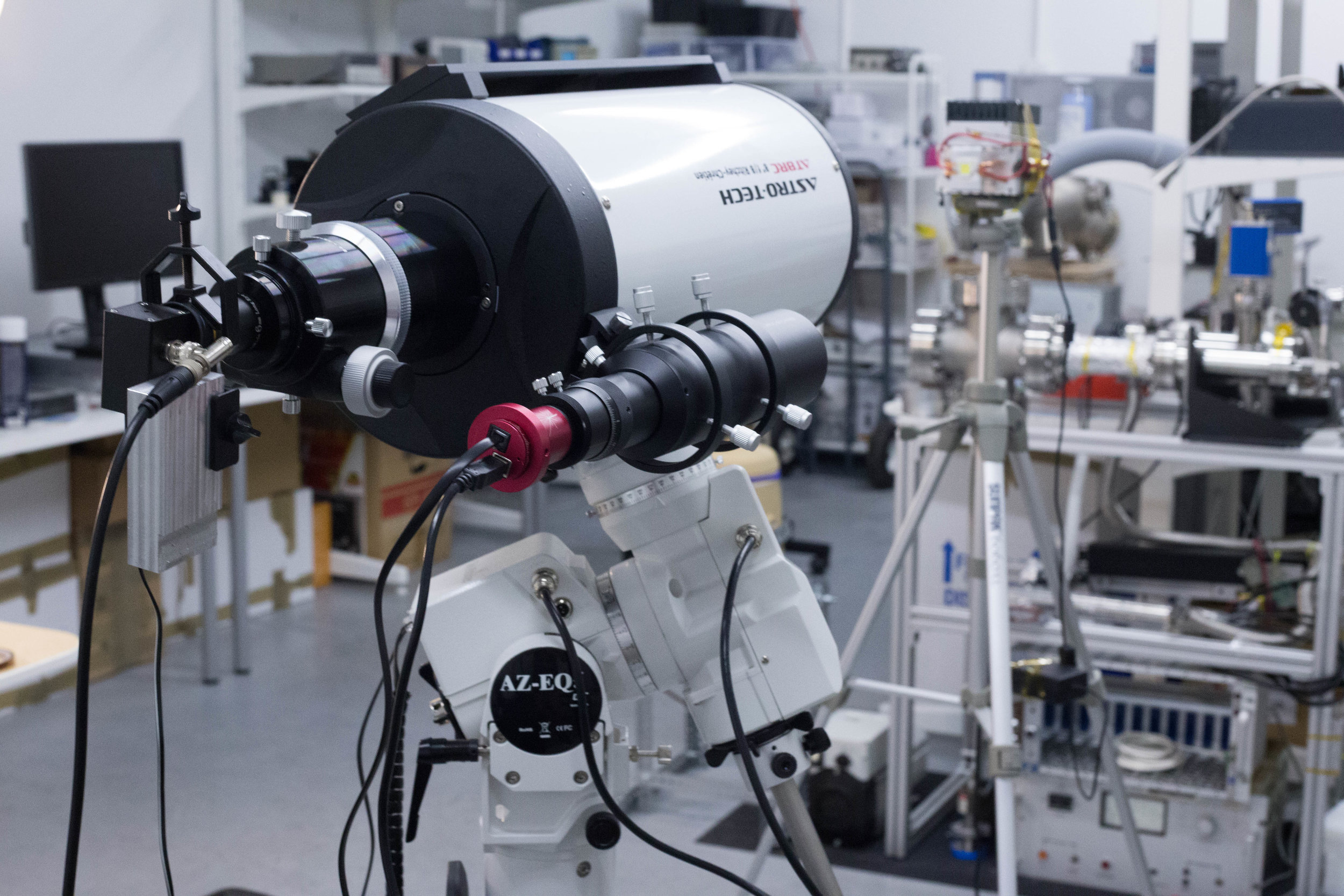}
\caption{Ground station prototype based on a Ritchey-Chretien telescope with coaxial tracking system and commercial mount}
\end{wrapfigure}

Each ground station is composed of a 30cm aperture Ritchey-Chretien telescope mounted on a commercial German Equatorial mount. Tracking is achieved via a coaxial 5cm refractor and a control computer. Both of the stations will be equipped with a YAG-pumped uplink laser modulated by a DAC/FPGA setup, which also allows the interference-based correction technique to be used on the PV-cell uplink. We expect the cost of each ground station in its final configuration to be under \$9000, a figure crucial for duplicability and widespread adoption to allow the construction of a ground station network. At the same time, the same hardware could be shared among multiple bands, together with laser downlink users, to spread deployment cost. A prototype is shown in Figure 9.

\section{Mission Goals}

The primary goal of the Calypso optical payload is to demonstrate the commercial viability of a fully optical spacecraft complete with an uplink and downlink. At the same time, a number of exciting technologies have presented themselves through the adoption of VLC. The mission also has the following goals.
\begin{itemize}
\item To measure atmospheric attenuation through power monitoring on the pre-calibrated transmitter
\item To train and evaluate a deep learning system for optical communication error correction
\item To evaluate the feasibility of differential phase delay correction using two ground stations
\end{itemize}

\section{Conclusion}
Building on a number of past missions directed at exploring the use of VLC on satellites, we have described a highly economical and space efficient bidirectional communication system designed for use on small satellites and as an emergency beacon. Utilizing state-of-the-art deep learning technology, we are able to implement a system in for use in optical communications that is capable of optimizing itself. This system is shown to have a number of unique advantages and can be easily commercialized based on present levels of maturity. Currently the flight version of the optical communication payload is under development and is expected to be integrated into the Cubesat by the end of 2017.

\bibliographystyle{plain}
\bibliography{bibtex_database.bib}







\end{document}